\documentclass[12pt]{article}

\usepackage{graphicx}
\usepackage{amsfonts}
\usepackage{mathptmx}
\usepackage{latexsym}
\usepackage[left]{lineno}
\usepackage{bbold}
\usepackage{bbm}

\begin{document}

%\linenumbers

\parskip 14pt
\baselineskip 0.2in

\title{Derivation of human chromatic discrimination ability from an information-theoretical notion of distance in color space}

%\subtitle{A notion of distance in color space}        % if too long for running head

\author{Mar\'{\i}a da Fonseca \and In\'es Samengo}

\date{Instituto Balseiro and Centro At\'omico Bariloche, (8400) San Carlos de Bariloche, Argentina.}

\maketitle

%================================================================

\begin{abstract}
The accuracy with which humans can detect small chromatic
differences varies throughout color space. For example, we are far
more precise when discriminating two similar orange stimuli than
two similar green stimuli. In order for two colors to be perceived
as different, the neurons representing chromatic information must
respond differently, and the difference must be larger than the
trial-to-trial variability of the response to each separate color.
Photoreceptors constitute the first stage in the processing of
color information; many more stages are required before humans can
consciously report whether two stimuli are perceived as
chromatically distinguishable or not. Therefore, although photoreceptor
absorption curves are expected to influence the accuracy of
conscious discriminability, there is no reason to believe that
they should suffice to explain it. Here we develop
information-theoretical tools based on the Fisher metric that
demonstrate that photoreceptor absorption properties
explain $\approx~87$\% of the variance of human color
discrimination ability, as tested by previous behavioral
experiments. In the context of this theory, the bottleneck in
chromatic information processing is determined by photoreceptor
absorption characteristics. Subsequent encoding stages modify only
marginally the chromatic discriminability at the photoreceptor
level.

%\keywords{Fisher information \and Color \and Cones \and
%Discrimination}
\end{abstract}

%================================================================

\section{Introduction}
\label{intro}

\begin{figure}[htdf]
\centerline{\includegraphics[keepaspectratio=true, clip = true,
scale = 1, angle = 0]{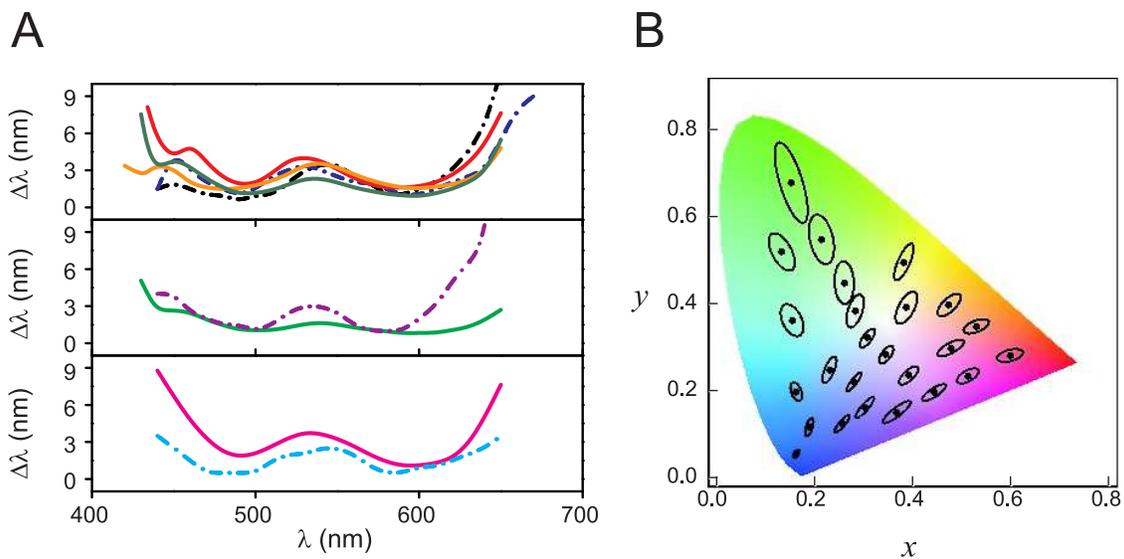}} \caption{\label{f1} {\bf Previous
experimental results of human color discrimination ability}. A:
Discrimination error $\Delta \lambda$ as a function of the
wavelength $\lambda$ of the reference beam, measured in behavioral
experiments for 9 different subjects. All subjects exhibit a local
maximum at $\lambda \approx 550$ nm. Subjects are separated in 3
groups, depending on the shape of the curve at $\lambda \approx
450$ nm. Top: Subjects exhibiting a local maximum. Middle:
Subjects exhibiting a shoulder. Bottom: Subjects with
monotonically decreasing errors. Data from Wright and Pitt (1934)
and Pokorny and Smith (1970). B: MacAdam ellipses (MacAdam 1942)
in the CIE 1931 $xy$ chromaticity diagram, reporting the region in
color space that is confounded with the center. As customary, each
ellipse is enlarged 10 times in each dimension for better visualization.  }
\end{figure}

Perception is the subjective experience that results from the
entire brain, not just photoreceptors. Color discrimination tasks
rely on the ability to detect small differences in the activity of
higher brain areas when two stimuli of similar chromatic
composition are presented. Human color discrimination ability has
been measured by several authors with behavioral experiments first
performed by Wright and Pitt (1934). In these studies, a bipartite
field was presented to a human subject. One half of the field,
here called the {\sl reference} field, was illuminated by a
monochromatic beam, constructed by filtering a broad-band light
source. The second half, the {\sl test} field, was also monochromatic,
and its wavelength was controlled by the observer. Initially, the
two beams had the same wavelength and luminosity. The observer was
instructed to displace the wavelength of the test field, until the
first noticeable difference in hue was perceived.  At this point,
the difference $\Delta \lambda$ between the reference and test
wavelengths was calculated. This difference constitutes the
discrimination error, that is, the interval in wavelengths below
which the two colors cannot be perceptually discriminated. The
discrimination error was reported to be a W-shaped function of
wavelength, as displayed in Fig.\ref{f1}A (Wright and Pitt 1934;
Pokorny and Smith, 1970).
Different curves correspond to different subjects. For all
subjects, discrimination errors were large towards the two borders
of the visible spectrum, and at around 550 nm, roughly at
the center of the visible spectrum where luminosity sensitivity is
maximal (Sharpe et al. 2005). At short wavelengths, the data show
some variability across subjects. We have therefore separated the
9 curves into 3 groups (displayed in different panels), depending
on whether the discrimination error exhibited an additional
maximum at approximately 450 nm (top panel), only a shoulder
(middle panel), or a monotonic behavior (bottom panel).

In 1942, David MacAdam (1942) broadened these discriminability
experiments testing the ability to distinguish any two neighboring
points in the entire color space, not just the subset of light
beams composed of a single wavelength. The concept of color, in
fact, cannot be restricted to wavelength. Blends of wavelengths
produce a new chromatic sensation that emerges exclusively from
the mixture, the hue of which differs from the hues of the
individual components. In order to test human chromatic
discrimination ability in the entire color space, MacAdam measured
ellipses in the CIE 1931 $xy$ chromaticity diagram, here shown in Fig.
1B. Points in this space represent hue and saturation, and are
independent of the total luminosity. Each ellipse in the diagram
indicates the area in color space (multiplied by 10, for better
 visualization) inside which two different stimuli
cannot be discriminated. 

In this study, we test how much of the results in Fig.~\ref{f1}
can be explained from a given noise model. The basic assumption is that
two colors can be discriminated when the trial-to-trial variability
of their neural representations is smaller than the difference of the 
corresponding means. The discriminability $d'$ of two colors is 
proportional to the square root of the Fisher information 
(Seung and Sompolinsky, 1993). The Fisher information, in turn, 
introduces a notion of distance in color space. Hence, our working 
hypothesis is that two colors become distinguishable when the Fisher 
distance between them is larger than a given fixed minimum: the detection 
threshold.  

Fisher Information has been successfully used as a tool to
disclose computational strategies in the nervous system for decades
(see for example Abbot and Dayan 1999, Dayan and Abbot 2001,
Brunel and Nadal 1998) and continues to be widely employed (Ganguli
and Simoncelli 2014, Wei and Stocker 2015).
Within the Fisher framework,
two previous studies (Clark and Skaff 2009; Zhaoping et al. 2011)
have derived the discrimination accuracy expected by an ideal
observer that only has access to the number of photons absorbed by
the three types of cones. Both studies were restricted to light
beams composed of a single wavelength, and succeeded in explaining
the W-shaped function of Fig.~\ref{f1}A. Our starting point is the work
of Zhaoping et al. (2011). We first express the main result of
their work in terms of an analytical expression for the
discrimination error. We use the theory to speculate how putative
tetrachromat subjects perceive the chromatic space. In the case of
trichromats, we interpret the subject-to-subject variations in
Fig.~\ref{f1}A as resulting from the reported variability in the
composition of the human retina. More importantly, we derive new
information-theoretical tools that expand the analysis
to the entire chromatic space, beyond monochromatic light beams.
By considering mixtures of wavelengths, we provide a theoretical
framework to also explain Fig.~\ref{f1}B. By focusing our
attention on the curved border of the chromatic space of Fig.~\ref{f1}B,
we also recover the previous result with monochromatic light
beams of Fig.~\ref{f1}A.

Our analysis concludes that 87\% of the variance of MacAdam's data
can be explained by properties of photoreceptors alone. 
Discrimination errors reported in behavioral experiments are the 
result of the entire chain of hue-dependent computations intervening 
in chromatic perception. Therefore, the agreement between experiment
and theory implies that 
the bottleneck in chromatic information
processing seems to be mainly determined by photoreceptor
activity. Subsequent encoding stages either
operate optimally or, if they do not, loose information in a
color-independent manner.

%================================================================

\section{Representations of color space}
\label{representations}

Color is the subjective sensation that results when a light beam
of spectrum $I(\lambda)$ impinges the eye. The set of all possible
spectra has infinite dimension, since for a continuum of
wavelengths $\lambda$ the intensity $I(\lambda)$ can vary
arbitrarily. The human visual system, however, is insensitive to
most of these dimensions. Classical behavioral color-matching
experiments demonstrated that for most observers, three
monochromatic light sources, conveniently mixed, suffice to
reproduce all visible colors. Therefore, the human visual system
projects the space of all possible spectra on a 3-dimensional
subspace. All spectra sharing the same projection are metamers,
that is, are perceived as indistinguishable. To represent colors
as 3-dimensional vectors, here we use the coordinates $(X, Y, Z)$
defined in the CIE 1931 (Appendix A1). When two color vectors
$(X_1, Y_1, Z_1)$ and $(X_2, Y_2, Z_2)$ only differ in their
length (they are proportional to one another), they share the same
hue and saturation, and can only be distinguished by their
luminosity. In some applications, it is desirable to discard the
luminosity dimension, and only retain the two remaining features.
The CIE 1931 meeting also established a convention to carry out
this reduction, by transforming the coordinates $(X, Y, Z)$ into a
new set of coordinates $(x, y, Y)$ defined by
\begin{equation} \begin{array}{lll}
x = \frac{X}{X + Y + Z}, & y = \frac{Y}{X + Y + Z}, & Y = Y. \end{array} \label{e15}
\end{equation}

\noindent 
The components $x$ and $y$ do not vary if $X, Y$ and $Z$ are all
multiplied by the same factor, so $x$ and $y$ no longer contain
the luminosity dimension. The variable $Y$ is associated with the
sensation of brightness, since for monochromatic spectra, the
wavelength dependence of $Y$ closely resembles the apparent
luminosity curve (Sharpe et al, 2005). MacAdam's experiment was
reported in CIE $xy$ chromatic space (Fig.~\ref{f1}).

\section{Statistics of the photon shower}
\label{shower}

\begin{figure}[htdf]
\centerline{\includegraphics[keepaspectratio=true, clip = true,
scale = 1, angle = 0]{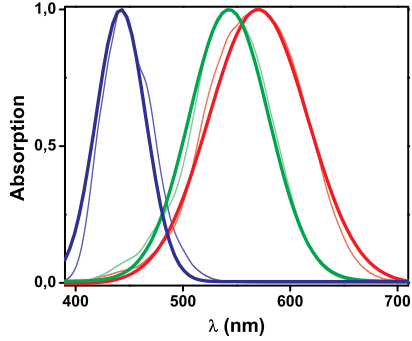}} \caption{\label{f4} Quantal cone
fundamentals for $S$ (blue), $M$ (green) and $L$ (red) cones, and
their analytic approximations. Thin lines: Data from Stockman and
Brainard (2009). Thick lines: fitted equation $h_i(\lambda) =
\exp[-(\lambda - \lambda_i)^2 / \sigma_i^2]$ with $\lambda_S =
442.1$  nm, $\lambda_M = 542.8$ nm, $\lambda_L = 568.2$  nm,
$\sigma_S = 32.96$ nm, $\sigma_M = 52.8$ nm, and $\sigma_L =
64.76$ nm. }
\end{figure}

For monochromatic light sources of mean intensity $I$ and
wavelength $\lambda$, the probability $P(\overrightarrow{K} |
\lambda)$ that $\overrightarrow{K} = (K_S, K_M, K_L)$ photons are
absorbed by $S, M$ and $L$ cones is (Appendix A2)
\begin{equation}
P[\overrightarrow{K} |\lambda]
= \prod_{i \in \{S, M, L\}} {\rm Poisson}[K_i | I q_i(\lambda)],
\label{e7}
\end{equation}
where ${\rm Poisson}(x | y)$ represents a Poisson distribution of
the random variable $x$ with mean $y$, $q_i(\lambda) = \beta_i
h_i(\lambda)$, the functions $h_i(\lambda)$ are the spectral
sensitivities of cones of type $i$ illustrated in Fig.~\ref{f4},
and the parameters $\beta_i$ are the normalized cross sections
associated with each fate of the photon. If $A_0$ is the retinal
area not covered by cones,
\begin{equation}
\beta_i = \frac{W_i A_i}{A_0 + \sum_{j \in \{S, M, L \}}
W_j A_j} \ \ {\rm for} \ i \in \{S, M, L\}, \ \ \
\beta_0 = \frac{A_0}{A_0 + \sum_{j \in \{S, M, L \}}
W_j A_j} \label{e1a}
\end{equation}
where $W_i$ is the number of photoreceptors of type $i$, and $A_i$
is the cross section of each cone.

The light source is now assumed to have an arbitrary spectrum
$I(\lambda)$. Defining the coefficients
\begin{equation}
\alpha_i = \int I(\lambda) \ q_i(\lambda) \ {\rm d}\lambda, \ \ \ {\rm for} \ i \in \{S, M, L\},
\label{e11}
\end{equation}
we show in Appendix A2 that
\begin{equation}
P[\overrightarrow{K} | I(\lambda)] =
\prod_{\ell \in \{S, M, L \}} \ {\rm Poisson}(K_i | \alpha_i).
\label{e12}
\end{equation}
As expected, Eq.~\ref{e12} reduces to Eq.~\ref{e7} when the
spectrum $I(\lambda)$ represents a monochromatic source, that is,
for $I(\lambda) = I \ \delta(\lambda - \lambda_0)$.

\section{The geometry of color space}
\label{geometry}

In simple discrimination experiments, the perceptual properties of
color are contained in the number of photons absorbed by $S$, $M$
and $L$ cones. The probability that $S$, $M$ or $L$ cones absorb
$K_S$, $K_M$ and $K_L$ photons, respectively, depends on the
properties of the impinging light beam. Here we consider two types
of experiments: Monochromatic beams of fixed intensity and varying
wavelength, and light sources composed of arbitrary spectra. In
the first case the light beam is characterized by the wavelength
$\lambda$, and in the second case, by the vector $(\alpha_S,
\alpha_M, \alpha_L)$. Distances in color space---in $\lambda$
space, or in $(\alpha_S, \alpha_M, \alpha_L)$ space---are defined
by the effect on the number of absorbed photons
$\overrightarrow{K}$ caused by changes in the composition of the
light source. The Fisher information $J$ is a metric tensor
(Appendix A3) that defines scalar products and distances in color
space. In the monochromatic case, since $\lambda$ is a
1-dimensional parameter, the Fisher tensor reduces to the scalar
\begin{equation}
J(\lambda) = - \left\langle \frac{\partial^2}{\partial \lambda^2}
\ln P \left( \overrightarrow{K} | \lambda \right)
\right\rangle_{P\left(\overrightarrow{K} | \lambda \right)},
\end{equation}
where the angular brackets indicate average with respect to the
indicated distribution. In the case of arbitrary mixtures of
wavelengths, $J$ is a tensor represented by a $3 \times 3$
matrix. In coordinates $ \overrightarrow{\alpha} = (\alpha_S,
\alpha_M, \alpha_L)$ its components are
\begin{equation}
J_{ij} = -\left\langle \frac{\partial^2}{\partial \alpha_i
\partial \alpha_j} \ln P \left( \overrightarrow{K} |
\overrightarrow{\alpha} \right) \right \rangle_{P
\left(\overrightarrow{K} |\overrightarrow{\alpha} \right)}. \label{e:fi}
\end{equation}

The Cram\'er-Rao bound relates the Fisher tensor to the accuracy
with which the random vector $\overrightarrow{K}$ can be used to
estimate the coordinates of color space. Formally, this means that
the mean quadratic error of any unbiased estimator of the
wavelength $\lambda$ or the coordinates $(\alpha_S, \alpha_M,
\alpha_L)$ from the absorbed photons $\overrightarrow{K}$ is
bounded from below (Appendix A3). 

\begin{figure}[htdf]
\centerline{\includegraphics[keepaspectratio=true, clip = true,
scale = 1, angle = 0]{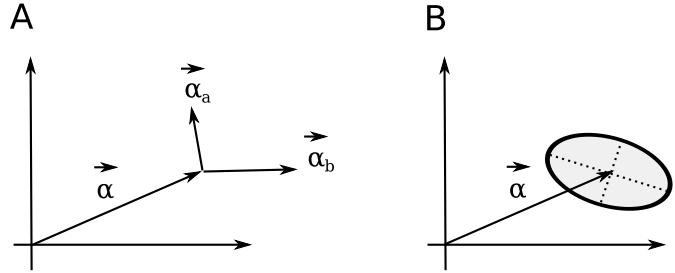}} \caption{\label{f3} {\bf Scalar
products in the space of parameters}. A: The Fisher information
metric tensor defines scalar products between two
vectors $\protect\overrightarrow{\alpha}^a$ and
$\protect\overrightarrow{\alpha}^b$ with common origin at location
$\protect\overrightarrow{\alpha}$ (Eq.~\ref{e0a}). B: The set
of all vectors at a constant distance from location
$\protect\overrightarrow{\alpha}$ defines an ellipsoid, whose
principal axes are displayed with dashed lines.}
\end{figure}

Metric tensors define scalar products. In Fig.~\ref{f3}A
we see how such products operate in the parameter space. The
scalar product between two vectors $\overrightarrow{\alpha}^a$ and
$\overrightarrow{\alpha}^b$ is
\begin{equation}
\langle \overrightarrow{\alpha}^a, \overrightarrow{\alpha}^b
\rangle = (\overrightarrow{\alpha}^a)^T \
J(\overrightarrow{\alpha}) \ \overrightarrow{\alpha}^b ,
\label{e0a}
\end{equation}
where the supra-script $T$ represents vector transposition. The
length of a vector $\overrightarrow{\alpha}^a$ originated at
$\overrightarrow{\alpha}$ is then
\begin{equation}
\left| \overrightarrow{\alpha}^a \right| = \sqrt{\langle
\overrightarrow{\alpha}^a, \overrightarrow{\alpha}^a \rangle} =
\sqrt{(\overrightarrow{\alpha}^a)^T \ J(\overrightarrow{\alpha}) \
\overrightarrow{\alpha}^a}. \label{e0b}
\end{equation}
Equation~\ref{e0b} implies that the set of vectors at a constant
distance of a certain $\overrightarrow{\alpha}$ is a conic. Since
the eigenvalues of the Fisher tensor are always non-negative, the
conic is an ellipsoid (Fig.~\ref{f3}B). The directions of the
principal axes of the ellipse are the eigenvectors of
$J(\overrightarrow{\alpha})$, which are also the eigenvectors of
$[J(\overrightarrow{\alpha})]^{-1}$. The lengths of those axes are
proportional to the inverse of the square root of the
corresponding eigenvalues of $J(\overrightarrow{\alpha})$, or
equivalently, to the square root of the eigenvalues of
$[J(\overrightarrow{\alpha})]^{-1}$. Since $J$ depends on
$\overrightarrow{\alpha}$, the size, excentricity and orientation
of the ellipse may well vary from point to point.

In Fig.~\ref{f1}B, the CIE 1931 $xy$ chromatic space has
coordinates $(x, y)$. Each ellipse measured by MacAdam represents
the set of points in color space where the first detectable
chromatic difference with the point at the center is perceived.
The ellipses represent the points at distance $\delta$ from the
center, where $\delta$ is the detection threshold. In this paper
we aim at evaluating up to which point the ellipses measured by
MacAdam can be derived from the properties of photoreceptors.

We now consider two different coordinate systems
$\overrightarrow{\alpha}$ and $\overrightarrow{\alpha}'$ to
represent colors. Let $\overrightarrow{F}$ be the vectorial
function involved in the mapping between them:
\begin{equation}
\overrightarrow{\alpha}' = \overrightarrow{F}(\overrightarrow{\alpha}). \label{e:newvariables}
\end{equation}
The matrix representation of the Fisher tensor transforms as (Appendix A3)
\begin{equation}
J(\overrightarrow{\alpha}) = C^T \ J'(\overrightarrow{\alpha}') \
C, \label{e:trans}
\end{equation}
with $C$ defined by the Jacobian matrix
\begin{equation}
C = \left(\begin{array}{ccc} \frac{\partial F_1}{\partial
\alpha_1} & \dots & \frac{\partial F_1}{\partial \alpha_d} \\
\vdots & & \vdots \\ \frac{\partial F_d}{\partial \alpha_1} &
\dots & \frac{\partial F_d}{\partial \alpha_d} \end{array}
\right). \label{e:c}
\end{equation}

%------------------------------------------------------------

\section{Discrimination of two similar wavelengths}
\label{puro}

For a monochromatic light beam of fixed intensity $I$,
Eq.~\ref{e7} establishes a probabilistic mapping between each
wavelength $\lambda$ and the vector $\overrightarrow{K}$. The
components $K_S, K_M$ and $K_L$ are converted to electrical
signals by photoreceptors, and then processed by the rest of the
brain. From the information-theoretic point of view, the data
processing inequality (Amari and Nagaoka, 2000) ensures that the
chromatic information encoded in later processing stages cannot
exceed the amount of chromatic information contained in
$\overrightarrow{K}$. A conscious subject, therefore, cannot have
better discrimination ability than that of an optimal estimator
inferring the wavelength $\lambda$ from photoreceptor activity,
that is, from $\overrightarrow{K}$. The optimal estimator is
usually referred to as the {\sl ideal observer}. The Cram\'er-Rao
bound in this case reduces to its 1-dimensional form (Cram\'er,
1946), $\Delta \lambda \ge 1 / \sqrt{J(\lambda)}$, implying that
the minimal error of the ideal observer is the inverse of the
square root of the Fisher information $J(\lambda)$.

Throughout the paper, spectral sensitivity curves were taken from
the cone fundamentals reported by Stockman and Brainard (2009). In
order to work with differentiable functions, the experimental
curves were approximated by functions $h_i(\lambda) =
\exp[-(\lambda -\lambda_i)^2/\sigma_i^2]$, with fitting parameters
$\lambda_i$ and $\sigma_i$, coinciding with the position of the
peak and the width of the data. Both the original and the fitted
curves are displayed in Fig.~\ref{f4} (fitted parameters in the
figure caption). Using this approximation, we insert Eq.~\ref{e7}
in Eq.~\ref{e:fi} and get
\begin{equation}
J(\lambda) = I \sum_{i \in \{S, M, L\}}
\frac{\left[q_i'(\lambda)\right]^2}{q_i(\lambda)} \label{e12a} \\
= 4 I\sum_{i \in \{S, M, L\}} \ \frac{(\lambda -
\lambda_i)^2}{\sigma_i^4} \ {\rm e}^{-(\lambda -
\lambda_i)^2/\sigma_i^2}. \label{e13}
\end{equation}
The formal expression of Eq.~\ref{e12a} was derived by Dayan and Abbot (2001),
and was first applied to the chromatic context by Zhaoping et al
(2011). Here we provide the analytical expression at the right of
Eq.~\ref{e13}. Note that the amount of Fisher information is
proportional to the intensity of the light source, in
Eq.~\ref{e13} represented by the mean number of photons $I$.

\begin{figure}[htdf]
\centerline{\includegraphics[keepaspectratio=true, clip = true,
scale = 1, angle = 0]{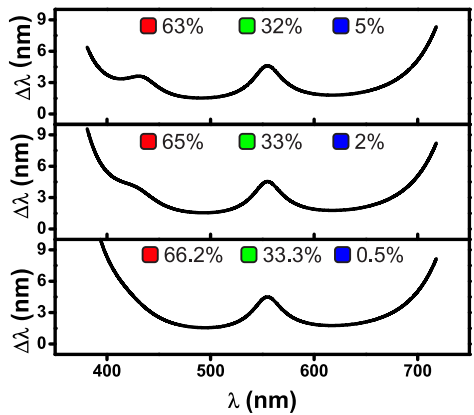}} \caption{\label{f6} {\bf Minimal
discrimination error $\Delta \lambda$ as a function of the
reference wavelength $\lambda$, obtained from Eq.~\ref{e13}}.
Different curves correspond to different
proportions of $S$ (blue), $M$ (green) and $L$ (red) cones, as
indicated in the legends. For all wavelengths, the mean
number of photons was taken as $I = 1000$.}
\end{figure}

In Fig.~\ref{f6}, we display the minimal estimation error $\Delta
\lambda = 1 / \sqrt{J(\lambda)}$
obtained with Eq.~\ref{e13}, for subjects whose retinas contain
different proportions of $S$, $M$ and $L$ cones. To draw the
figure, we set the mean number of photons to $1000$, in order to
match the experimental conditions, where weak photopic
illumination was employed. The shapes of the theoretical curves
are qualitatively similar to the ones measured experimentally
(Fig.~\ref{f1}A).

Previous studies have shown that there is substantial
subject-to-subject variation in the proportions of different types
of cones (Hofer et al. 2005). The variability in cone distribution
suffices to explain the different types of behavioral results.
Specifically, a shoulder appears in the short-wavelength region
only for subjects whose proportion of $S$ cones exceeds 2\%,
whereas a full local maximum requires $\beta_S \ge 5\%$. The
larger the proportion of $S$-cones, the higher the peak at $\sim
450$ nm.

The variability in the proportion of $S$-cones is the crucial
factor determining the shape of $\Delta\lambda$. Humans also
display a remarkable variability in the relative proportion of $M$
and $L$ cones (Roorda and Williams 1999), involved in
Eq.~\ref{e13} through the factors $\beta_M$ and $\beta_L$.
However, as long as the total amount $\beta_M + \beta_L$ remains
constant, relative variations do not modify the shape of the curve.
The cone fundamentals of $M$ and $L$ cones are close
to each other, so varying the relative proportion $\beta_M  /
\beta_L$ produces a negligible effect in $\Delta \lambda$.

\begin{figure}[htdf]
\centerline{\includegraphics[keepaspectratio=true, clip = true,
scale = 1, angle = 0]{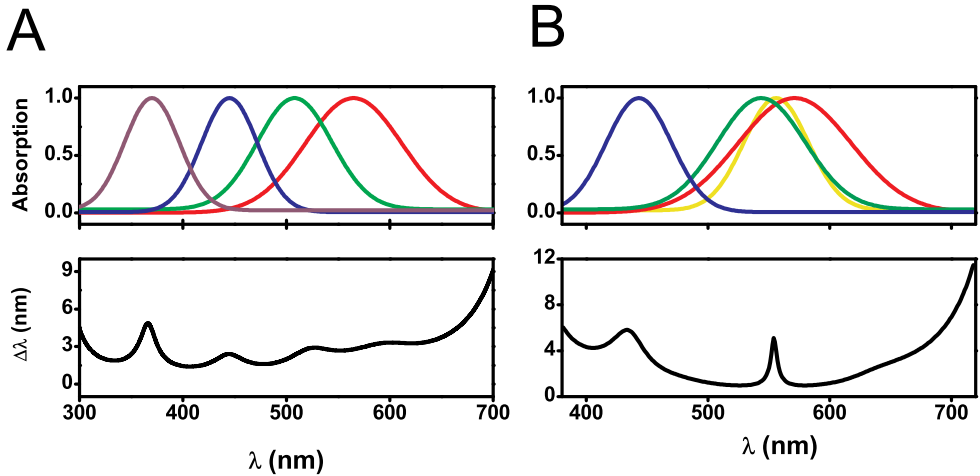}} \caption{\label{f7} {\bf Minimal
discrimination error $\Delta \lambda$ as a function of wavelength
$\lambda$, obtained from a generalization of Eq.~\ref{e13} that
considers 4 different types of cones}. A. Top: Absorption curves
of estrilid finch (Hart et al. 2000) with local maxima at
wavelengths $\lambda_i = 368, 445, 508, 565$ nm, and widths
$\sigma_i = 38.58, 38.44, 52.87, 65.8$ nm, respectively. Bottom:
Predicted discrimination error, with mean number of photons set to
1000. Parameters $\beta_i$ are set to 0.25 for all cones. B. Top:
Putative cone fundamentals of a human tetrachromat (Jordan et al.
2010). The additional curve (in yellow) peaks at $\lambda_y = 555$
nm, and has width $\sigma_y = 38.51$ nm. Bottom: Predicted
discrimination error, with mean number of photons set to 1000.
Parameters $\beta_i$ are set to 5\%, 31.33\%, 31.34\% and 31.33\%
for $S, M, Y$, and $L$ cones, respectively.}
\end{figure}

The theoretical framework developed here can also be used to
predict the wavelength dependence of discrimination in observers
that are not available for experimentation, either for their
rarity, or for their non-human nature. Figure~\ref{f7} displays
the minimal discrimination errors in fortunate subjects endowed
with 4 different types of cones. In panel A, bird vision is
discussed. Absorption curves are approximately
equidistant from each other (Hart et al. 2000) giving rise to
accurate color discrimination abilities that extend further into
the ultraviolet spectrum.  Four local maxima are visible in
$\Delta \lambda$, accounting for each of the 4 absorption curves.

In panel B, we display the results for a putative human
tetrachromat, for which the extra cone is hypothesized to lie
between the $M$ and $L$ cones, as anomalous subject cDa29 studied
by Jordan et al. (2010). In spite of the incorporation of an
additional curve, the discrimination ability of this subject is
similar to that of normal trichromats. The substantial overlap
between $M$ and $L$ absorption curves of trichromats implies that the
addition of one more absorption curve in the same wavelength
region makes virtually no difference. This does not
mean that the putative tetrachromat of Fig.~\ref{f7}B perceives
the same color space as trichromats, since the present discussion
is restricted to the discrimination of neighboring monochromatic
beams. Color space also includes mixtures of wavelengths (see
below), and some mixtures, for example the purples obtained by
mixing red and blue, are not metameric with any single
monochromatic beam. Tetrachromats may perceive many more mixtures
that cannot be mapped on the trichromat color space. Our analysis
predicts, however, that their ability to discriminate neighboring
monochromatic beams remains essentially unaltered.

%----------------------------------------------------------------

\subsection{Discrimination of spectra composed of mixtures of wavelengths}
\label{mezcla}

\begin{figure}[htdf]
\centerline{\includegraphics[keepaspectratio=true, clip = true,
scale = 1, angle = 0]{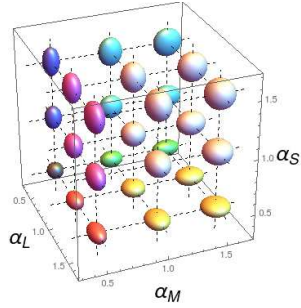}} \caption{\label{f8}{\bf Ellipsoids
indicating the regions of space
$\protect\overrightarrow{\protect\alpha}$  lying at a fixed
distance of each central point}.}
\end{figure}

To extend the previous analysis to the entire color space, the
Fisher information should be written as a function of coordinates
that describe the chromatic composition of an arbitrary beam
$I(\lambda)$. Equation \ref{e12} implies that the probability
distribution of the absorbed photons $\overrightarrow{K}$ is
blind to all aspects of the spectrum $I(\lambda)$ not contained in
the vector $\overrightarrow{\alpha}$ defined in Eq.~\ref{e11}.
Replacing
Eq.~\ref{e12} in \ref{e:fi}, we get
\begin{equation}
J \left(\overrightarrow{\alpha} \right) = \left(\begin{array}{ccc}
\frac{1}{\alpha_S} & 0 & 0 \\ 0 & \frac{1}{\alpha_M} & 0 \\
0 & 0 & \frac{1}{\alpha_L} \end{array} \right). \label{e14}
\end{equation}
The metric tensor is diagonal, so the ellipsoids
defining the points at constant distance of a given parameter
$\overrightarrow{\alpha}$ have their principal axes aligned with
the coordinate axes. The square root of the inverse of
$J\left(\overrightarrow\alpha\right)$ defines an ellipsoid
around each color $\overrightarrow{\alpha}$ where
all points are at the same distance from the central point
$\overrightarrow{\alpha}$ (Fig.~\ref{f8}).

In order to compare with experimental data, we need to transform
the metric tensor of Eq.~\ref{e14} from the parameter space
$\overrightarrow{\alpha}$ to the CIE 1931 $xy$ chromatic space
where MacAdam reported the minimal discriminable ellipses. We
perform the transformation in two steps. First, we change from
$(\alpha_S, \alpha_M, \alpha_L)$ to $(X, Y, Z)$, and then from
$(X, Y, Z)$ to the triplet $(x, y, Y)$. So far, the two
transformations are invertible. Once we have the Fisher matrix in
the space $xyY$, we take the submatrix associated to the
components $xy$ alone, in order to compare with MacAdam's
experiment.

Each of the two transformations involves a $C$-matrix defined in
Eq.~\ref{e:c}. If we call $C_1$ the matrix of the first
transformation, and $C_2$ the one of the second, the two
concatenated transformations are implemented by a matrix $C = C_1
C_2$. To calculate $C_1$ we analyze the way the color matching
functions transform, when passing from $(\alpha_S,$ $\alpha_M,
\alpha_L)$ to $(X, Y, Z)$. By fitting a linear transformation
between the two, we deduce that
\[
\left(\begin{array}{c} \alpha_S \\ \alpha_M \\ \alpha_L \end{array} \right) = C_1
 \ \left(\begin{array}{c} X \\ Y \\ Z \end{array} \right), \ \ \ \ \ \ \ \
{\rm with} \ \ \ \ 
C_1 = \left( \begin{array}{ccc} 0.038 \beta_S & -0.043
\beta_S & 0.48 \beta_S \\ -0.39\beta_M & 1.17 \beta_M & 0.049 \beta_M \\
0.34 \beta_L & 0.69 \beta_L & -0.076 \beta_L
\end{array} \right). \]
To calculate $C_2$, we invert Eq.~\ref{e15} and find
\[ X = Y \ x / y \ \ \ \ \ \ \ \ \ \ \ \  Y = Y \ \ \ \ \ \ \ \ \ \ \ \
Z = Y \ (1 - x - y) / y.
\]
Using Eq.~\ref{e:c}, we obtain
\[
C_2 = \left(\begin{array}{ccccc} \frac{Y}{y} & & -\frac{x Y}{y^2}
& &  \frac{x}{y} \\ & & & & \\ 0 & & 0 & & 1 \\ & & & & \\
-\frac{Y}{y} & & - \frac{Y(1 - x)}{y^2} & & \frac{1 - x - y}{y} \end{array}
\right).
\]
With the resulting matrix $C = C_1 C_2$, we calculate the Fisher tensor in 
space $xyY$,  and then focus on the submatrix
corresponding to the first two components. All the coefficients of
the obtained $2 \times 2$ submatrix are proportional to the
luminosity variable $Y$. Hence, the lengths of the principal axes
of the ellipses defining the equidistant colors are proportional to $Y^{-1/2}$, and
the area is proportional to $1/Y$. Other than this scaling factor,
the luminosity variable has no additional effect. Since all other
variables appearing in the Fisher tensor are adimensional, the
units with which we measure distances in the $xy$ space are
$[Y]^{-1/2}$. Here we use MacAdam's unit of color difference
(Wyszecki and Stiles 2000), implying that the distance between
each central point and the ellipse measured by MacAdam is unity.
In this system, the coordinate $Y$ is adimensional.

\begin{figure}[htdf]
\centerline{\includegraphics[keepaspectratio=true, clip = true,
scale = 1, angle = 0]{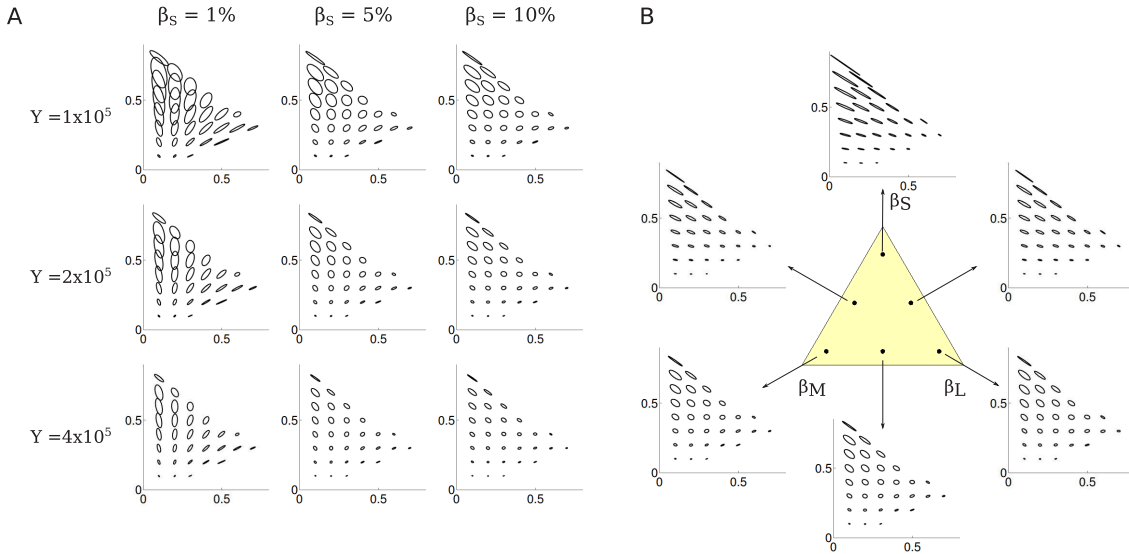}} \caption{\label{f9} {\bf Ellipses
obtained by transforming the ellipsoids in Fig.~\ref{f8}}. $xy$
CIE chromatic coordinates represented by the horizontal and vertical
axes, respectively. A:
Dependence of ellipses on parameters $\beta_S$ and $Y$. We set
$\beta_M = \beta_L = (1 - \beta_S)/2$. B: Dependence of ellipses
on retinal composition, for $Y = 2 \times 10^5$. The triangle represents the
accessible area of the space $(\beta_S, \beta_M, \beta_L)$.
Indicated panels correspond to
$\protect\overrightarrow{\protect{\beta}} = (\beta_S, \beta_M,
\beta_L) = (0.8, 0.1, 0.1)$ (top), $(0.45, 0.1, 0.45), (0.1, 0.1,
0.8), (0.1, 0.45, 0.45), (0.1, 0.8, 0.1)$ and $(0.45, 0.45, 0.1)$
as we rotate clockwise. As customary, ellipses are enlarged 10
times in each dimension for better visualization. }
\end{figure}

In Fig.~\ref{f9} the ellipses at distance 1 from 31 center points are
displayed.
In A, $\beta_S$ is varied within the physiological range. As
$\beta_S$ increases (from left to right) the ellipses become 
smaller, and more compressed along the direction $(1, -1)$. 
Increasing the value of $Y$ (from top to bottom)
shrinks the ellipses.

In B, we display the ellipses for several retinal compositions,
without restricting the values of $\beta_i$ to the realistic
range. In this context, any set of $(\beta_S, \beta_M, \beta_L)$
defines a possible retina, as long as all $\beta_i$ are positive,
and the three of them sum up to unity. In the space of possible
$\overrightarrow{\beta}$ vectors, these conditions define the
triangle illustrated in Fig.~\ref{f9}B. As we move along the
bottom border of the triangle, we confirm that the relative
proportion of $\beta_M$ and $\beta_L$ does not change the ellipses
qualitatively (three bottom panels). Increasing $\beta_S$, instead
(moving upward) reduces the size of the ellipses in the direction
$(1, 1)$, and augments them along the direction $(-1, 1)$. In
other words, increasing the proportion of $\beta_S$ helps
discriminating blue vs. yellow stimuli, but has a detrimental
effect on the discrimination of red vs. green. The ellipses
corresponding to the inner area of the triangle smoothly
interpolate those at the border.

In order to compare with MacAdam's experiment, we need to fit the
parameters $\beta_S$ and $Y$, both kept fixed during the
experiment. To do so, we systematically vary $\beta_S \in (0,
0.1)$ and $Y \in (0, 10^6)$ and compare the theoretical ellipses
evaluated at the 25 points measured by MacAdam with the 25
experimental ellipses. The optimal parameters are the ones that
make both sets of ellipses maximally similar. To do so, we need a
criterion of similarity between ellipses. Two concentric ellipses
may differ in their size, their orientation, or their
excentricity. In order to evaluate the three aspects
simultaneously, and to adequately weigh the relevance of each, we
define the distance between two concentric ellipses as the
Kullback-Leibler divergence between two Gaussian distributions
whose covariance matrices are defined by the tested ellipses. As
the two distributions become more and more similar, the two
ellipses merge into one another, implying a simultaneous match
between size, elongation and excentricity. Averaging over the 25
measured points, $\beta_S$ and $Y$ are fitted by minimizing
\[
D =\frac{1}{25} \sum_{i = 1}^{25} {\rm D}_{KL} \left[{\cal N}({\bf r}_i,
\Sigma_i^{th})||{\cal N}({\bf r}_i, \Sigma_i^e)\right],
\]
where the sum runs over the 25 colors tested by MacAdam, ${\rm
D}_{KL}$ is the Kullback-Leibler divergence, ${\cal N}({\bf r}_i,$ $
\Sigma)$ is a normal bivariate distribution centered at the colors
$\bf{r}_i$ where MacAdam performed his experiment, and with
covariance matrix $\Sigma$. The supra-index $th$ represents the
theoretical matrix, and $e$ the experimental one. The experimental
covariance matrix is constructed from the reported ellipses: We
calculate the matrix whose eigenvectors are in the directions of
the principal axes reported by MacAdam, and whose eigenvalues
coincide with the lengths of the principal axes. The theoretical
covariance matrix is the inverse of the Fisher information. An
analytical form for the Kullback-Leibler divergence for
multivariate Gaussian distributions is derived in Duchi (2014).
When $D$ is employed as a fitting criterion, the goodness-of-fit
may be defined in terms of an $R^2$-value defined as $R^2 = 1 -
D/{\rm D}_e$, where ${\rm D}_e$ is the average Kullback-Leibler
divergence between all experimental ellipses.

\begin{figure}[htdf]
\centerline{\includegraphics[keepaspectratio=true, clip = true,
scale = 0.6, angle = 0]{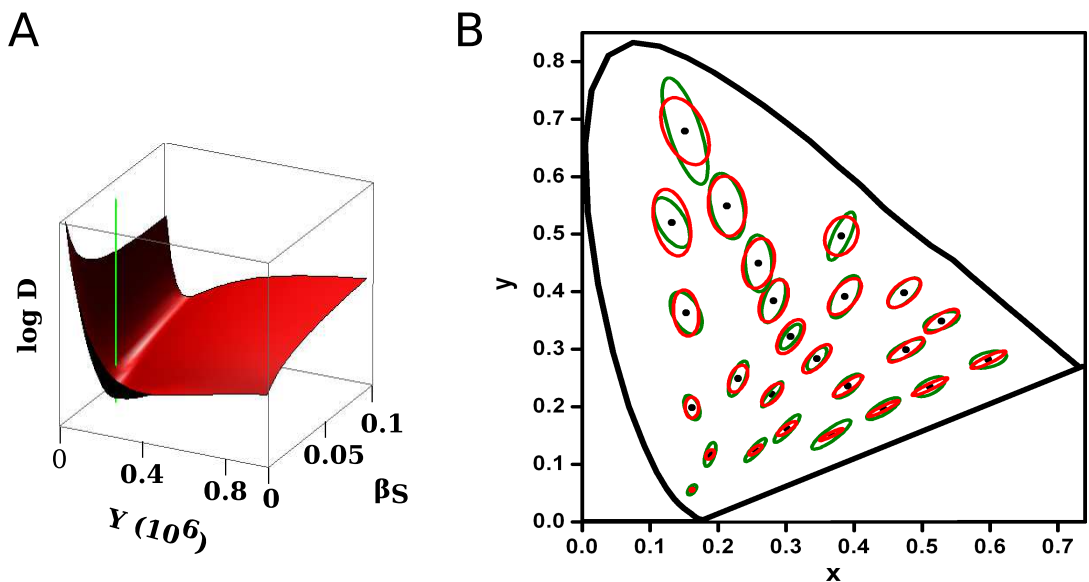}} \caption{\label{f10} {\bf
Comparison between theory and experiment}. A: Distance $D$
between theoretical and experimental ellipses as a function of
parameters $\beta_S$ and $Y$. The green vertical line indicates
the location of the optimal parameter values $\beta_S = 2.1\%$ and $Y =
184,000$. B: Ellipses measured by MacAdam (green) compared to the
ones derived from our theoretical model (red) for the optimal
parameters. As customary, each ellipse is enlarged 10 times in
each dimension for
better visualization. }
\end{figure}

In Fig.~\ref{f10}A we see the dependence of $D$ with parameters
$\beta_S$ and $Y$. The optimal values are $\beta_S = 2.1\%$ and $Y
= 184,000$, for which $D = 0.36$, and $R^2 = 0.87$.
In Fig.~\ref{f10}B, we see the model is effective in describing
the variation of the size, orientation and excentricity of the
ellipses throughout the chromatic space. More
quantitatively, the obtained $R^2$ value implies that the theory
explains 87\% of the variability of the experimental data.

%================================================================

\section{Discussion}

Here we derived a metric in color space from a noise model of the
representation of color in the brain. Several classical studies
have derived the minimal discrimination ellipses from line
elements (Wyszecki and Stiles, 2000). Those theories used
heuristic arguments to propose a distance in color space. Not
being framed in the Fisher geometry, they do not entail a
data processing inequality nor a Cram\'er-Rao bound. The advantage
of the Fisher metric is that it brings along a rigorous
mathematical framework, first, for deriving the metric from a
noise model, and then, for transforming the metric from one space
to another. Given the noise model, the Fisher metric is
undisputable. Of course, there are many candidate noise models,
depending on which neuronal processes are described.
The confrontation of the derived Fisher ellipses with experimental
data actually provides a systematic way to evaluate the adequacy
of alternative noise models.

Here we offer an attempt to perform such confrontation, using one
particular noise model based on the sole description of the photon
absorption process. We conclude that a simple Poisson model of the
statistics of photoreceptor absorption account for $\approx 87$\%
of the variance of the behavioral results in the $xy$ chromatic
space. Our theory also predicts that the minimal discrimination
error is inversely proportional to the square root of the light
intensity, following a Rose-DeVries law, originally reported in
contrast discrimination thresholds at low light intensities (Rose
1948, DeVries 1943), and later confirmed for chromatic
discrimination experiments (Rovamo et al 2001). Experiments show
that this dependence holds in the low-intensity photopic regime,
but loses validity as the light intensity becomes larger (Rovamo
et al 2001). Therefore, additional optic or neural color-dependent
processing stages not contained in the Poisson photoreceptor model
must come into play at high intensities.

In spite of having neglected all subsequent color processing
stages beyond absorption, even the voltage variations in the inner
segments of photoreceptors, the distances derived from the Fisher
approach reproduce a large fraction of the experimental
variability. The derived ellipses are only guaranteed to coincide
with the measured discrimination error when the Cram\'er-Rao bound
is tight, that is, when further processing stages perform
optimally, or at least, they do not introduce additional
color-dependent distortions. A priori, there is no reason to
believe that such should be the case. The similarity between the
theoretical and the experimental results therefore suggests that
photon absorption constitutes the crucial stage in the chromatic
dependence of color processing ability, and it suffices to explain
most of the structure observed in experimental data. All
subsequent processing stages either perform optimally or, if they
lose information, they do so in a color-independent manner.

The Cram\'er-Rao bound of Eq.~\ref{e:cramer} is only valid for
unbiased estimators, a more complex formula is required in the
biased case (Cover and Thomas, 1991). However, in the presence of
achromatic backgrounds (as in all experiments explored here), 
discrimination errors have been always reported to
have zero  mean, so we work under the assumption that the nervous
system is able to implement at least one unbiased estimator, for
which Eq.~\ref{e:cramer} holds.
Different is the case where the target and test stimuli are
presented against a chromatic background, where subjects have been
reported to bias their estimation of the target stimulus away from
the hue of  the background (see for example Klaue and Wachtler
2015). In such cases, we suspect that the
more complex form of the Cram\'er-Rao bound should be employed.

The Fisher tensor determines the distance between neighboring
colors; the distance between distant colors must be calculated
by adding the infinitesimal distances encountered along a 
specific path. If one of the two colors lies at a border of the
chromatic space, the path must be entirely contained inside
the space. When the path connecting two colors is short, all the 
involved infinitesimal distances are obtained from essentially 
the same Fisher tensor, since the Fisher metric varies smoothly 
with location. When comparing the theoretical and experimental 
ellipses, we have assumed that the Fisher distance between the 
central color and the ellipse could be calculated with a single 
Fisher tensor: the one of the center. To assess the validity of 
the approximation, we verified that inside each ellipse the 
eigenvalues of the theoretical Fisher tensor vary at most 1.27 
\%, and the inclination angle at most $0.97^\circ$  (recall that 
all depicted ellipses have been enlarged 10 times in each 
dimension, for better visibility).

We are aware of two other previous studies where chromatic
discrimination ability was modeled by information-theoretical
methods. The first one (Clark and Skaff, 2009) was based on
stronger assumptions as the ones used here, since the properties
of chromatic perception were explained in terms of a specific
decoding process that takes place (explicitly or implicitly) in
the visual system. Our work neither supports nor refutes the
proposed decoding, we simply show that it is not strictly required
to explain a large fraction of the variance in human
discrimination ability. The second study was developed by Zhaoping
et al. (2011). Their approach was the starting point for the
present study. They also discussed how the Fisher metric varies
with mean light intensity. Instead here, we have focused on (a)
providing an analytical formula for the monochromatic case, (b)
extending the analysis to the whole chromatic space, and (c)
discussing the discrimination ability of observers endowed with
different retinal compositions.

Chromatic discrimination ability is limited by the imprecision
with which neighboring colors are represented in the brain. The
stochasticity considered here regards the unpredictability of the
exact proportion of photons captured by $S$, $M$ and $L$ cones,
given that the three absorption curves overlap with each other.
The variance of $K_i$ is $K q_i (1 - q_i)$, and is maximal when
both $q_i$ and $1 - q_i$ are far from zero. For $M$ and $L$ cones,
this condition is met at approximately 550 {\sl nm}, where both
absorption probabilities are high. Human color discrimination
error has a local maximum at $\approx$ 550 {\sl nm}, roughly
coinciding with the wavelength where humans perceive maximal
luminosity (Sharpe et al. 2005). So far, this coincidence appeared
as incidental. An analysis of the equations involved in our study,
however, reveals that color discrimination ability is determined
by the derivative of the quantal cone fundamentals: The larger the
derivative, the larger the value of the Fisher information (Dayan
and Abbott, 2001). Since $L$ and $M$ cone fundamentals are very
similar, and given that the variance is particularly large at
$\approx$ 550 {\sl nm}, the two maxima cannot be separated apart,
and discrimination error peaks at a wavelength that is
approximately the average of the wavelengths where $L$ and $M$
absorption curves reach their maxima. The coincidence, hence, is
grounded on the mathematical properties of the Fisher information.
If the number of $S$ cones is large enough, a local maximum in
discrimination error is also achieved at the wavelength where the
S-cone absorption curve peaks, $\approx$ 450 {\sl nm}. Moreover,
the theory also predicts how discrimination ability varies with
retinal composition, suggesting that the variability in anatomical
properties of different observers may account for the variability
in the experimental data.

Throughout our work, we have only considered cones, although rod
absorption is also modulated by wavelength. By a
simple extension of our analysis, it is also possible to include
rods in the evaluation of chromatic discrimination ability.
However, since rods and cones have different luminosity
sensitivity, the comparison with behavioral data should be
performed with experiments where the total light intensity was
controlled. The parameters $\beta_i$ scaling the relevance of each
photoreceptor should also include a factor accounting for the
different cross sections of rods and cones, and their differential
sensitivity depending on the total luminosity. The analysis
presented here is only valid for photopic illumination conditions
(as reported by the experiments) where rods are assumed to be
saturated. Extensions to other models, including rods or other
optical and neural processes, are possible. Results can be
expressed in the classical color spaces employed here, through the
transformation formulas of Sect.~\ref{geometry}.

\section*{Appendix}

\subsection*{A1: Transformations of the light spectrum leading to representations of color}

\begin{figure}[htdf]
\centerline{\includegraphics[keepaspectratio=true, clip = true,
scale = 1, angle = 0]{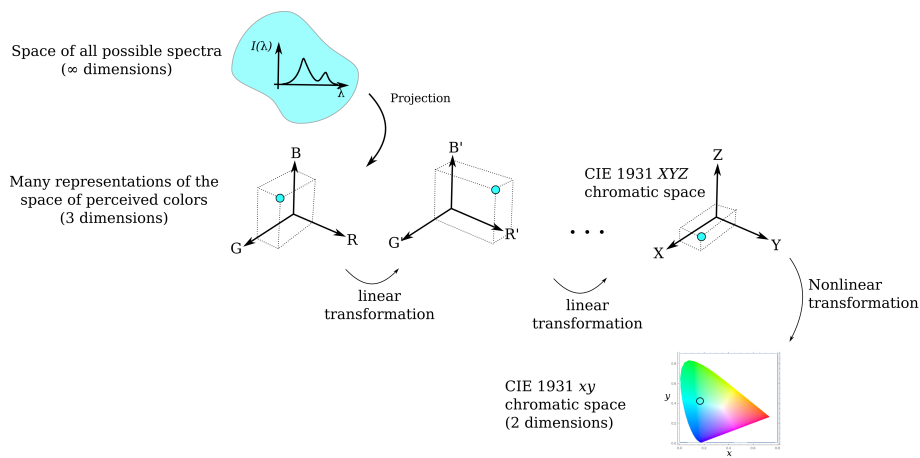}} \caption{\label{f2} {\bf
Transformations between different representations of the
composition of a light beam}. Top: The most complete
representation is the spectrum $I(\lambda)$, specifying the energy
density in each wavelength. The space of all possible spectra has
infinite dimensions. Middle: The human visual system can only
perceive 3 dimensions. The projection from the space of spectra to
the space of chromatic perceptions is linear (Grassmann's law).
There are many representations of the 3-dimensional space
perceived by humans. One of them is the CIE 1931 $XYZ$ color space
(middle bottom). Each spectrum at the top projects to a single
point in the three-dimensional space by means of a non-invertible
transformation. Bottom: The CIE 1931 $xy$ chromatic space is a
nonlinear transformation of the CIE 1931 $XYZ$ space that
eliminates the luminosity dimension, and only keeps variations in
hue and saturation. Each point of the three-dimensional space maps
onto a point in the $xy$ space.}
\end{figure}

From the physical point of view, the spectrum $I(\lambda)$
provides a complete characterization of a light beam. The space of
all possible spectra has infinite dimensions. Color matching
experiments performed by Helmholtz and Young proved that by
adjusting the intensity of three monochromatic sources of fixed
wavelengths, human trichromats construct a beam that they perceive
as visually indistinguishable from a target light source of
arbitrary spectrum. Hence, the human visual system projects the
high-dimensional space of all possible spectra onto three
dimensions (Fig.~\ref{f2}).
When the target light source is monochromatic and has wavelength
$\lambda$, the three intensities required to construct the mixture
define the color matching functions $\bar{b}(\lambda),
\bar{g}(\lambda)$ and $\bar{r}(\lambda)$, whose functional shape
depends on the wavelengths of the three sources than comprise the
mixture. The matching operation is linear, implying that the
visual appearance of an arbitrary spectrum $I(\lambda)$ is
governed by three numbers, defined as
\begin{equation}
\begin{array}{lll}
B = \int I(\lambda)\bar{b}(\lambda) \ {\rm d}\lambda,
&
G = \int I(\lambda)\bar{g}(\lambda) \ {\rm d}\lambda,
&
R = \int I(\lambda)\bar{r}(\lambda) \ {\rm d}\lambda . \end{array} \label{e.0}
\end{equation}
If the wavelengths of the three fixed light sources are varied, the
shape of the color matching functions changes. Different 
$(R, G, B)$ representations have thus appeared, depending on
the chosen wavelengths. In fact, any
invertible linear transformation of one set of coordinates $(R, G,
B)$ yields a new set of coordinates $(R', G', B')$ equally
valid, represented in Fig.~\ref{f2} as one of the coordinate
systems in the middle column. The new coordinates can also be
obtained from integrals like Eq.~\ref{e.0}, but with new functions
$\bar{b}'(\lambda), \bar{g}'(\lambda), \bar{r}'(\lambda)$, derived
from a linear transformation of the old functions. In 1931, the
International Commission on Illumination (CIE, for its initial in
French) selected a particular set of coordinates $(X, Y, Z)$,
associated with specific color matching functions usually notated
as $\bar{x}(\lambda), \bar{y}(\lambda), \bar{z}(\lambda)$ (Wyszecki and Stiles, 2000).

\subsection*{A2: Poisson absorption models}

In this appendix, we derive Eqs.~\ref{e7} and \ref{e12}. Repeated
use is made of the formulas
\[ {\rm{\bf Binomial:}}  \ \ (a + b)^n = \sum_{j = 0}^n \frac{n!}
{j! (n - j)!} \ a^j \ b^{n - j}; \ \ \ \ {\rm{\bf Multinomial:}}  \ \
\left( \sum_{i = 1}^k a_i \right)^n = n! \ \sum_{j_1, \dots, j_k}
\prod_{\ell = 1}^k \frac{a_\ell^{j_\ell}}{j_\ell!},
\]
where the sum of the multinomial theorem runs over all sets of
integers $\{j_1, \dots, j_k \}$ fulfilling the conditions $0 \le
j_\ell \le n$ and $n = j_1 + \dots + j_k$.

\subsubsection*{Monochromatic light source of fixed intensity}

When a photon of wavelength $\lambda$ impinges on the retina under
central photopic illumination conditions, four outcomes are
possible: The photon may be detected by a cone of type $S$, $M$ or
$L$, or it may pass undetected. The probability of each outcome
depends on the fraction of $S$, $M$ and $L$ cones that tile the
retina and on the probability that each cone absorbs a photon of
wavelength $\lambda$, also called the {\sl spectral sensitivity}
of each cone. Once these parameters are known, from the
statistical point of view, illuminating the retina with $I_0$
photons of wavelength $\lambda$ is equivalent to randomly
distributing $I_0$ balls into 4 boxes whose cross sections depend
on the wavelength $\lambda$. The probability that $N_S, N_M$ and
$N_L$ fall on $S, M$ and $L$-cones respectively, and that $N_0$ do
not fall on cones is
\begin{equation} \label{e1}
P(\overrightarrow{N}|I_0) = I_0! \prod_{i \in \{S, M, L, 0\}}
\frac{\beta_i^{N_i}}{N_i !},
\end{equation}
where $\overrightarrow{N} = (N_S, N_M, N_L)$. The components $N_S,
N_M, N_L, N_0$ are not all independent, since they must sum up to
$I_0$. Therefore, $N_0$ is a shorthand notation for $N_0 = I_0 -
N_S  - N_M - N_L$.

\begin{figure}[htdf]
\centerline{\includegraphics[keepaspectratio=true, clip = true,
scale = 0.7, angle = 0]{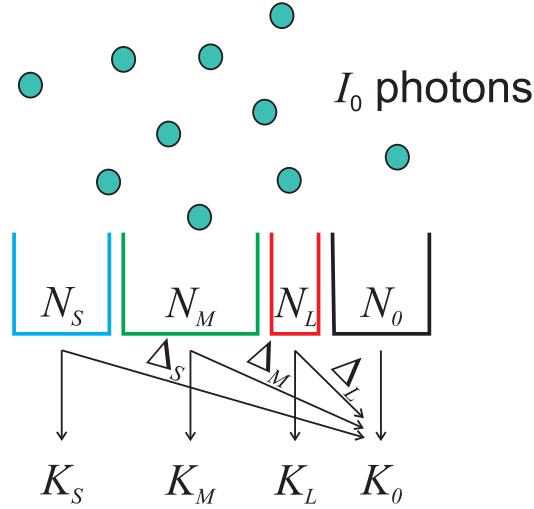}} \caption{\label{f5} $I_0$ photons
of wavelength $\lambda$ fall on $S$, $M$ or $L$ cones (colored
boxes), or miss cones altogether (black box). Each of the $N_i$
photons that fall on cones of type $i$ may either be absorbed (and
become one of the $K_i$ photons absorbed by a photoreceptor of
type $i$), or pass unabsorbed (and become one of the $\Delta_i$
photons not absorbed by a cone of type $i$). The total number of
photons absorbed by a cone of type $i$ is $K_i = N_i - \Delta_i$,
and the number of photons that remain unabsorbed is $K_0 = N_0 +
\Delta_S + \Delta_M + \Delta_L$. The process that transforms the
$K$ impinging photons into $(N_S, N_M, N_L, N_0)$ is governed by
the multinomial distribution of Eq.~\ref{e1}, and the one that
transforms $N_i$ into $K_i$, by the binomial of Eq.~\ref{e2}. }
\end{figure}

When $N_i$ photons reach a cone of type $i$, the probability that
$K_i$ of them are absorbed is a binomial distribution with
absorption probability $h_i(\lambda)$ (Fig.\ref{f4}),
\begin{equation} \label{e2}
P(K_i|N_i) = \frac{N_i!}{K_i! (N_i - K_i)!} \ h_i(\lambda)^{K_i} \
[1 - h_i(\lambda)]^{N_i - K_i}, \ \ \ {\rm for \ \ } i \in \{S, M, L\}.
\end{equation}
When these two processes are coupled sequentially, the fate of
each photon is decided through the processes depicted in
Fig.~\ref{f5}.
The probability that $K_S, K_M, K_L$ photons are absorbed by cones
$S, M, L$ and that $K_0$ pass undetected is
\begin{equation}
\label{e3} P(\overrightarrow{K}|I_0, \lambda) = \sum_{{\bf N} > {\bf K}}
P(\overrightarrow{N} | I_0, \lambda ) \ \prod_{i \in \{S, M,
L\}} \ P(K_i | N_i),
\end{equation}
where the sum ranges over all vectors $\overrightarrow{N} = (N_S,
N_M, N_L)$ that fulfill the conditions $N_S \ge K_S, N_M \ge K_M,
N_L \ge K_L$ and $N_S + N_M + N_L \le I_0$. Replacing
Eqs.~\ref{e1} and \ref{e2} in Eq.~\ref{e3}, defining the
scaled absorption probabilities $q_i(\lambda) = \beta_i
h_i(\lambda)$, for $i \in \{S, M, L\}$, and $q_0(\lambda) = 1 -
q_S(\lambda) - q_M(\lambda) - q_L(\lambda)$ and the 
the numbers $\Delta_S, \Delta_M$ and $\Delta_L$ of lost photons 
(see Fig.~\ref{f5})
\[
\begin{array}{ccc}
\Delta_S = N_S - K_S, &
\Delta_M = N_M - K_M, &
\Delta_L = N_L - K_L, \end{array}
\]
we get, after some algebraic manipulations,
\begin{equation} \label{e5}
P(\overrightarrow{K} |I_0, \lambda) = I_0! \  \prod_{i \in \{S, M, L,
0\}} \frac{q_i(\lambda)^{K_i}}{K_i!}.
\end{equation}
\normalsize The derivation involved the use of the binomial
theorem three times. The composition of the multinomial process of
Eq.~\ref{e1} and the binomial of Eq.~\ref{e2} yields another
multinomial distribution governed by the scaled absorption
probabilities, combining the parameters governing the two
processes in play. In Eq.~\ref{e5}, the variables $K_S, K_M$ and
$K_L$ are not independent, since the distribution also includes a
factor that depends on $K_0 = I_0 - K_S - K_M - K_L$.

\subsubsection*{Monochromatic light source of variable intensity}

If the total number of photons $I_0$ of wavelength $\lambda$ is a
stochastic variable governed by a Poisson distribution of mean
$I$
\begin{equation} \label{e6}
P[I_0 | I] = {\rm e}^{- I} \ \frac{I^{I_0}}{I_0!},
\end{equation}
then the probability of the absorbed photons is
\begin{equation} \label{e6a}
P[\overrightarrow{K} |\lambda, I] = \sum_{I_0 = 0}^{+\infty}
P(\overrightarrow{K} |I_0, \lambda) \ P[I_0 | I].
\end{equation}
Replacing Eq.~\ref{e5} and \ref{e6} in Eq.~\ref{e6a}, and after
some algebraic manipulations, we arrive at Eq.~\ref{e7}. A light
source with variable intensity, hence, gives rise to absorbed
photon counts $K_S, K_M, K_L$ that are independent from one
another.

\subsubsection*{Light sources of arbitrary spectrum}

We now consider a light source composed of photons of $r$
different wavelengths $\lambda_j$, where $j$ ranges between $1$
and $r$. The mean number of photons of the different wavelengths
defines an $r$-dimensional vector $\overrightarrow{I} =
(I(\lambda_1), \dots, I(\lambda_r))$. There are many ways in
which $S$ cones can absorb $K_S$ photons: All the $K_S$ photons
may have the same wavelength $\lambda_1$, half of them may have
wavelength $\lambda_1$ and the other half wavelength $\lambda_2$,
etc. Here we consider all the possibilities. We define $G_S^j$ as
the number of photons of wavelength $\lambda_j$ absorbed by cones
$S$, and arrange these numbers in $r$-dimensional vectors
\[\begin{array}{ccc}
{\bf G}_S = (G_S^1, \dots, G_S^r), &
{\bf G}_M = (G_M^1, \dots, G_M^r), &
{\bf G}_L = (G_L^1, \dots, G_L^r). \end{array}
\]
If the total numbers of absorbed photons are $K_S, K_M$ and $K_L$,
the components of the three vectors defined above must sum up to
these values, that is,
\begin{equation}
\begin{array}{ccc}
\sum_{j = 1}^r G_S^j = K_S, &
\sum_{j = 1}^r G_M^j = K_M, &
\sum_{j = 1}^r G_L^j = K_L. \end{array} \label{e8}
\end{equation}
We call ${\cal U}_S$, ${\cal U}_M$ and ${\cal U}_L$ the sets of
all vectors ${\bf G}_S$, ${\bf G}_M$ and ${\bf G}_L$ whose
components are non-negative integers fulfilling Eqs.~\ref{e8}.
Mathematically, for $i \in \{S, M, L\}$,
\[
{\cal U}_i = \{{\bf G}_i / G_i^j \ge 0 \ \forall j \ \& \
\sum_{j = 1}^r G_i^j =
K_i \},
\]
The probability of cones $S, M$ and $L$ of absorbing
$\overrightarrow{K}$ photons can be written in terms of the sum of
all possible spectral compositions of the absorbed photons,
namely,
\begin{equation}
P(\overrightarrow{K}| \vec{I} \ ) = \sum_{{\bf G}_S \in {\cal U}_S}\sum_{{\bf G}_M \in {\cal
U}_M}\sum_{{\bf G}_L \in {\cal U}_L} \prod_{j = 1}^r P[G_S^j,
G_M^j, G_L^j | I(\lambda_j)], \label{e9}
\end{equation}
where the probability $P[G_S^j, G_M^j, G_L^j | I(\lambda_j)]$ in
the right-hand side of Eq.~\ref{e9} is the same as the one of
Eq.~\ref{e7}, but is now evaluated in a $G$-vector (as opposed to
a $K$-vector). The sums represent the fact that many combination
of wavelengths may contribute to the same $\overrightarrow{K}$.

Replacing Eq.~\ref{e7} in Eq.~\ref{e9}, and using the
multinomial theorem, we get
\begin{equation}
P(\overrightarrow{K} | \vec{I})
= \prod_{\ell \in \{S, M, L\}}\frac{\exp\left[-\sum_{i = 1}^r
I(\lambda_i)q_\ell(\lambda_i)\right]}{K_\ell!} \  \left[\sum_{j = 1}^r
I(\lambda_j)q_\ell(\lambda_j)\right]^{K_\ell}. \label{e10}
\end{equation}
Defining the coordinates $(\alpha_S, \alpha_M, \alpha_L)$
\begin{equation}
\alpha_i = \sum_{j = 1}^r I(\lambda_j)q_i(\lambda_j), \ \ \ {\rm
for} \ i \in \{S, M, L\}, \label{e11a}
\end{equation}
Eq.~\ref{e10} yields Eq.~\ref{e12}. If the spectrum contains a
continuum of wavelengths $\lambda$ with mean spectral energy
$I(\lambda)$, the calculations performed here remain unchanged,
except for the fact that the coefficients $\alpha_i$ must be
defined in terms of integrals (compare Eq.~\ref{e11} and
\ref{e11a}).

\subsection*{A3: The Fisher geometry}

Here, the properties of the light beam that are relevant to color
discrimination are represented by a vector of parameters
$\overrightarrow{\alpha} = (\alpha_1, \dots, \alpha_d)$. In this
appendix, the parameter $\overrightarrow{\alpha}$ is not
necessarily defined by Eqs.~\ref{e11}. For example, when applied 
to experiments performed with monochromatic beams, we can take 
$d = 1$, and $\overrightarrow{\alpha}$ equal to the wavelength
$\lambda$ (or equivalently, any invertible function of the
wavelength). In the case of mixtures, we may take $d = 3$, and
$\overrightarrow{\alpha}$ defined by Eqs.~\ref{e11}, or equivalently,
as $\overrightarrow{\alpha} = (X, Y, Z)$. We may also consider 
$d = 2$ and $\overrightarrow{\alpha} = (x, y)$.

If the probability distribution of a random variable
$\overrightarrow{K}$ depends on the parameter
$\overrightarrow{\alpha}$, a notion of distance can be defined in
the $\overrightarrow{\alpha}$ space, quantifying the effect of
changing $\overrightarrow{\alpha}$ on $P(\overrightarrow{K} |
\overrightarrow{\alpha})$. It may well be the case that in certain
regions of the $\overrightarrow{\alpha}$ space, a displacement of
the parameter in a certain amount ${\rm d}\overrightarrow{\alpha}$
changes the distribution $P(\overrightarrow{K} |
\overrightarrow{\alpha})$ radically, whereas in other regions the
same displacement hardly has an effect. In these circumstances,
distances in the $\overrightarrow{\alpha}$ space vary from point
to point: The same displacement ${\rm d}\overrightarrow{\alpha}$
corresponds to a large distance in the first case, and to small
one in the second. The Fisher information introduced in Eq.~\ref{e:fi} 
defines a metric tensor that gives rise to a notion of distance
in parameter space: the length of a vector is given by Eq.~\ref{e0b},
and the distance between two neighboring vectors $\overrightarrow{\alpha}^a$ 
and $\overrightarrow{\alpha}^{b}$ is the length of $\overrightarrow{\alpha}^a - \overrightarrow{\alpha}^{b}$. 
The Cram\'er-Rao bound relates the Fisher tensor to the accuracy
with which the random variable $\overrightarrow{K}$ can be used to
estimate the parameter $\overrightarrow{\alpha}$. If the Fisher
information is large, sampling $\overrightarrow{K}$ can provide a
good estimate of $\overrightarrow{\alpha}$, if an efficient
decoding procedure is used. A low Fisher information, in contrast,
implies that $P(\overrightarrow{K}|\overrightarrow{\alpha})$
hardly varies with $\overrightarrow{\alpha}$ and therefore, it is
impossible to make (on average) a good guess of the value of
$\overrightarrow{\alpha}$ by sampling $\overrightarrow{K}$, not
even with an optimal decoding procedure. Formally, this means that
the mean quadratic error of any unbiased estimator
$\hat{\overrightarrow{\alpha}}(\overrightarrow{K})$ of the
parameter $\overrightarrow{\alpha}$ is bounded from below. We
define the mean quadratic error as a $d\times d$ matrix $E$, with
elements
\[
E_{ij}(\overrightarrow{\alpha}) =  \left\langle \left[
\hat{\alpha}_i(\overrightarrow{K}) - \alpha_i\right] \left[
\hat{\alpha}_j(x) - \alpha_j\right] \right\rangle_{P
\left(\overrightarrow{K}|\overrightarrow{\alpha}\right)}.
\]
The Cram\'er-Rao bound states that
\begin{equation}
E(\overrightarrow{\alpha})\cdot J(\overrightarrow{\alpha}) \ge
\mathbbm{1}, \label{e:cramer}
\end{equation}
where $\mathbbm{1}$ is the identity matrix, and the inequality
implies that all the eigenvalues of the matrix $E\cdot J$ cannot
be smaller than unity. The Cram\'er-Rao bound of Eq.~\ref{e:cramer}
can also be expressed as $E \ge J^{-1}$. Therefore, $J^{-1}$ is
the minimal mean quadratic estimation error. The
larger the information, the smaller the error, and vice versa. The
bound expressed in Eq.~\ref{e:cramer} is only valid for unbiased
estimators, a more complex formula is required in the biased case
(Cover and Thomas, 1991).

In Sect.~\ref{representations}, several representations of the
composition of a light beam were introduced. One may, for example,
represent the light beam with the spectrum $I(\lambda)$, or with
specific coordinates $RGB$, or with the CIE 1931 $XYZ$, or the
reduced $xy$. Assume that new coordinates
$\overrightarrow{\alpha}'$ are defined from old coordinates
$\overrightarrow{\alpha}$ by means of a transformation
$\overrightarrow{F}$ (see Eq.~\ref{e:newvariables}). If the
elements of the Fisher information matrix for the representation
$\overrightarrow{\alpha}$ are known, one may calculate their value
in the representation $\overrightarrow{\alpha'}$. The
transformation must be such as to preserve scalar products. If
$\overrightarrow{\alpha}^a$ and $\overrightarrow{\alpha}^b$ are
two infinitesimal displacements from the vector
$\overrightarrow{\alpha}$, the transformed infinitesimal
displacements $\overrightarrow{\alpha}'^a$ and
$\overrightarrow{\alpha}'^b$ are defined from the first order
expansion of $\overrightarrow{F}$,
\begin{eqnarray}
\overrightarrow{F}(\overrightarrow{\alpha} +
\overrightarrow{\alpha}^a) &\approx& \overrightarrow{F}
(\overrightarrow{\alpha}) + \left[(\overrightarrow{\alpha}^a)^T
\overrightarrow{\nabla} \right] \overrightarrow{F}
(\overrightarrow{\alpha}) \equiv \overrightarrow{\alpha}' +
\overrightarrow{\alpha}'^a  \nonumber \\
\overrightarrow{F}(\overrightarrow{\alpha} +
\overrightarrow{\alpha}^b) &\approx&
\overrightarrow{F}(\overrightarrow{\alpha}) + \left[
(\overrightarrow{\alpha}^b)^T \overrightarrow{\nabla} \right]
\overrightarrow{F}(\overrightarrow{\alpha}) \equiv
\overrightarrow{\alpha}' + \overrightarrow{\alpha}'^b  \nonumber
\end{eqnarray}
Therefore, $\overrightarrow{\alpha}'^i = \left[
(\overrightarrow{\alpha}^i)^T \overrightarrow{\nabla} \right]
\overrightarrow{F}(\overrightarrow{\alpha})$, for $i \in \{a,
b\}$. Preserving the scalar product means that
\[
(\overrightarrow{\alpha}^a)^T \ J(\overrightarrow{\alpha}) \
\overrightarrow{\alpha}^b = (\overrightarrow{\alpha}'^a)^T \
J'(\overrightarrow{\alpha}') \ \overrightarrow{\alpha}'^b.
\]
Since the displacements $\overrightarrow{\alpha}^a$ and
$\overrightarrow{\alpha}^b$ are arbitrary, we arrive at
Eqs.~\ref{e:trans} and \ref{e:c}. In general, the matrix $C$
depends on the parameter $\overrightarrow{\alpha}$. Only if
$\overrightarrow{F}$ is a linear transformation,  $C$ reduces to a
constant matrix.

%\begin{figure}
%  \includegraphics{example.eps}
%\end{figure}
%\begin{figure*}
%  \includegraphics[width=0.75\textwidth]{example.eps}
%\end{figure*}
%\begin{table}
%\begin{tabular}{lll}
%\hline\noalign{\smallskip}
%\end{tabular}
%\end{table}

\section*{Acknowledgements}
We thank Rodrigo Echeveste for very productive discussions.

\section*{References}
\begin{enumerate}

\item[-] Abbot, L. F., \& Dayan, P. (1999). The effect of
correlated variability on the accuracy of a population
code. Neural Computation 11, 91-101.

\item[-] Amari, S.-I., \& Nagaoka, H. (2000). Methods in
information Geometry. Providence RI: American Mathematical
Society.

\item[-] Brunel N., \& Nadal J. P. (1998).
Mutual information, Fisher information and population coding.
Neural Computation 10(7), 1731-1757.

\item[-] Clark, J.J., \& Skaff, S. (2009). A spectral theory of
color perception. Journal of the Optical Society of America
26(12),  2488-2502.

\item[-] Cover, T.M., \& Thomas J. A. (1991). Elements of Information
Theory. New York: Wiley.

\item[-] Cram\'er, H. (1946). A contribution to the theory of
statistical estimation, Scandinavian Actuarial Journal 1946(1),
458-463.

\item[-] Dayan, P., \& Abbot, L. F. (2001). Theoretical
Neuroscience. Computational and Mathematical Modeling of
Neural Systems. Cambridge: MIT Press.

\item[-] DeVries, H. L. (1943). The quantum character of light and its
bearing upon threshold of vision, the differential sensitivity and
the visual acuity of the eye. Physica 10: 553-564.

\item[-] Duchi, J. C. (2014). Derivations for Linear Algebra and
Optimization. Resource document.
http://ai. stanford.edu/\~jduchi/projects/general\_notes.pdf.
Accessed Jan 2016.

\item[-] Ganguli, D. \& Simoncelli, E. P., (2014). Efficient Sensory Encoding
and Bayesian Inference with Heterogeneous Neural Populations.
Neural Computation 26(19),  2103-2134.

\item[-] Hart, N. S., Partridge, J. C., Bennet, A. T. D., \& Cuthill,
I.C. (2000). Visual pigments, cone oil droplets and ocular media
in four species of estrildid finch. Journal of Comparative
Physiology A 186, 681-694.

\item[-] Hofer, H., Carroll, J., Neitz, J., Neitz, M., \&
Williams, D. R. (2005). Organization of the Human Trichromatic Cone
Mosaic. Journal of Neuroscience, 25(42), 9669-9679.

\item[-] Jordan, G., Deeb, S. S., Bosten, J. M., \& Mollon, J. D.
(2010). The dimensionality of color vision in carriers of
anomalous trichromacy. Journal of Vision 8(10), 1-19.

\item[-] Klaue, S., \& Wachtler. T. (2015). Tilt in color space:
Hue changes induced by chromatic surrounds. Journal of Vision
15(13): 17, 111.

\item[-] MacAdam, D. L. (1942). Visual Sensitivities to Color
Differences in Daylight. Journal of the Optical Society of America
32(5), 247-274.

\item[-] Pokorny, J., \& Smith, V. C. (1970). Wavelength
discrimination in the presence of added chromatic fields. Journal
of the Optical Society of America 60(4), 562-569.

\item[-] Roorda, A., \& Williams, D. R. (1999). The arrangement of
the three cone classes in the living human eye. Nature 397,
520-522.

\item[-] Rose, A. (1948) The sensitivity performance of the human
eye on an absolute scale. Journal of the Optical Society of
America 28:196-208.

\item[-] Rovamo, J. M., Kankaanp\"a\"a, M. I., \& Hallikainen, J. (2001)
Spatial neural modulation transfer function for human foveal
visual system for equiluminous chromatic gratings. Vision Research
41:1659-1667.

\item[-] Sharpe, L. T., Stockman, A., Jagla, W., \& J\"agle, H.
(2005). A luminous efficiency function $V^*(\lambda)$ for daylight
adaptation. Journal of Vision 5, 948-968.

\item[-] Seung, H. S., \& Sompolinsky, H. (1993). Simple models
for reading neuronal population codes. PNAS USA 90: 10749-10753.

\item[-] Stockman, A., \& Brainard, D. H. (2009). Color vision
mechanisms, in Bass, M. (ed.) OSA Handbook of Optics, New York:
McGraw-Hill.

\item[-] von Helmholtz, H. (1896). Handbuch der physiologischen
Optik. Leipzig: Voss.

\item[-] Wei, X. X., \& Stocker, A. A. (2015). A Bayesian observer
model constrained by efficient coding can explain 'anti-Bayesian'
percepts. Nature Neuroscience 18, 1509-1517.

\item[-] Wright, W. D., \& Pitt, F. H. G. (1934). Hue discrimination
in normal colour vision. Proceedings of the Physical Society 46,
459-473.

\item[-] Wyszecki, G., \& Stiles, W. S. (2000). Color Science:
Concepts and Methods, Quantitative Data and Formulae. New York:
Wiley Interscience.

\item[-] Zhaoping, L., Geisler, W. S., \& May, K. A. (2011). Human
Wavelength Discrimination of Monochromatic Light Explained by
Optimal Wavelength Decoding of Light of Unknown Intensity. PLoS
ONE 6(5), e19248. doi:10.1371/journal.pone. 0019248.

\end{enumerate}

\end{document}